\documentstyle[12pt]{article}
\newcommand\be{\begin{equation}}
\newcommand\ee{\end{equation}}
\newcommand\bea{\begin{eqnarray}}
\newcommand\eea{\end{eqnarray}}
\newcommand\ket[1]{|#1\rangle}
\newcommand\bra[1]{\langle #1|}

\newcommand{\fatalpha}{{\bf \alpha \kern -0.44em \alpha}}
\newcommand{\fatsigma}{{\bf \sigma \kern -0.54em \sigma}}
\newcommand{\tpchi}{{\bf \chi \kern -0.35em \chi}}
\newcommand{\llambda}{{\bf \lambda \kern -0.45em \lambda}}

\bibliography{plain}
\title{\bf Exact calculation of robustness of entanglement via convex semi-definite programming} \vspace{20mm}
\author{ M. A. Jafarizadeh$^{a,b,c}$
 \thanks{E-mail:jafarizadeh@tabrizu.ac.ir},
M. Mirzaee$^{a,b}$ \thanks{E-mail:mirzaee@tabrizu.ac.ir}, M.
Rezaee$^{a,b}$ \thanks{E-mail:karamaty@tabrizu.ac.ir}
\\
\\
$^a${\small Department of Theoretical Physics and Astrophysics,
Tabriz University, Tabriz 51664, Iran.} \\ $^b${\small Institute
for Studies in Theoretical Physics and Mathematics, Tehran
19395-1795, Iran.} \\ $^c${\small Research Institute for
Fundamental Sciences, Tabriz 51664, Iran.}} \pagebreak

\vspace{20mm}
\begin{document}
\maketitle \vspace{15mm}
\newpage
\begin{abstract}
In general the calculation of robustness of entanglement for the
mixed  entangled quantum states is rather difficult to handle
analytically. Using the the convex  semi-definite programming
method,  the robustness of entanglement of  some mixed  entangled
quantum states such as: $2\otimes 2$ Bell decomposable (BD)
states, a generic two qubit state in Wootters basis,
iso-concurrence decomposable states, $2\otimes 3$ Bell
decomposable states, $d\otimes d$ Werner and isotropic states, a
one parameter $3\otimes 3$ state and finally multi partite
isotropic state, is calculated exactly, where thus obtained
results are in agreement with those of :$2\otimes 2$ density
matrices, already calculated by one of the authors in
\cite{Bell1,Rob3}. Also an analytic expression is given for
separable states that wipe out all entanglement and it is further
shown that they are on the boundary of separable states as
pointed out in \cite{du}.

{\bf PACs Index: 03.65.Ud }
\end{abstract}
\vspace{70mm}
\newpage
\section{INTRODUCTION}
Quantum entanglement has recently been attracted much attention as
a potential resource for communication and information processing
\cite{ben1,ben2}. Entanglement  usually arises from quantum
correlations between separated subsystems which can not be created
by local actions on each subsystem. By definition, a mixed state
$\rho$ of a bipartite system is said to be separable (non
entangled) if it can be written as a convex combination of product
states
$$
\rho=\sum_{i}w_{i}\,\rho_i^{(1)}\otimes\rho_i^{(2)},\qquad w_i\geq
0, \quad \sum_{i}w_i=1,
$$
where $\rho_i^{(1)}$ and $\rho_i^{(2)}$ are states of subsystems
$1$ and $2$, respectively. Although, in the case of pure states of
bipartite systems it is easy to check whether a given state is, or
is not entangled, the question is yet an open problem in the case
of mixed states. Therefore having a measure to quantify
entanglement of mixed states is likely to be valuable and several
measures of entanglement have been proposed
\cite{ben3,ved1,ved2,woot}.

One useful quantity introduced  in \cite{vidal} as a measure of
entanglement is robustness of entanglement. It corresponds to the
minimal amount of mixing with locally prepared states which washes
out all entanglement.  An analytical expression for pure states of
bipartite  systems by using partial transpose  has been given  in
\cite{vidal}. Meanwhile the robustness of entanglement has been
calculated for a Werner states. Moreover, in \cite{vidwer} Vidal
and Werner have computed the robustness of entanglement for
density operators with symmetry.  In \cite{rudolph} Rudolph,
using cross norm has clarified the relationship of the greatest
cross norm with the robustness of entanglement and has determined
the value of the greatest cross norm for Bell diagonal states. A
geometrical interpretation of robustness is given in \cite{du}
and it is pointed out  that two corresponding separable states
needed to wipe out all entanglement are necessarily on the
boundary of separable set.  On the other hand,  the robustness of
entanglement of  few mixed quantum states such as: $2\otimes 2$
Bell decomposable (BD) states and a generic two qubit state in
Wootters basis is already calculated by one of the authors
in\cite{Bell1, Rob3}. In Ref \cite{Rajagopal} has characterized
the robustness of entanglement, and its relation to the
permutation symmetries, for the basic set of eight entangled
three particle states of spin-1/2 objects. Authors in
\cite{Simon}, have studied  the robustness of multi-party
entanglement under local decoherence, modeled by partially
depolarizing channels acting independently on each subsystem.
Unfortunately, in general, the above mentioned quantity as the
most proposed measures of entanglement involves exteremization
which is difficult to handel analytically.

On the other hand, over the past years, semidefinite programming
(SDP) has been recognized as valuable numerical tools for control
system analysis and design. In (SDP) one minimizes a linear
function subject to the constraint that an affine combination of
symmetric matrices is positive semidefinite. SDP, has been studied
(under various names) as far back as the 1940s. Subsequent
research in semidefinite programming during the 1990s was driven
by applications in combinatorial optimization\cite{Luo04},
communications and signal processing \cite{Luo03,Luo02,Luo01},
and other areas of engineering\cite{Luo05}. Although semidefinite
programming is designed to be applied in numerical methods it can
be used for analytic computations, too. Some authors try to use
the SDP to construct an explicit entanglement witness
\cite{Doherty,Parrilo}. Kitaev used semidefinite programming
duality to prove the impossibility of quantum coin flipping
\cite{13}, and Rains gave bounds on distillable entanglement using
semidefinite programming \cite{17}. In the context of quantum
computation, Barnum, Saks and Szegedy reformulated quantum query
complexity in terms of a semidefinite program \cite{1}. The
problem of finding the optimal measurement to distinguish between
a set of quantum states was first formulated as a semidefinite
program in 1972 by Holevo4, who gave optimality conditions
equivalent to the complementary slackness conditions \cite{10}.
Recently, Eldar, Megretski and Verghese showed that the optimal
measurements can be found efficiently by solving the dual
followed by the use of linear programming \cite{3}. Also in
\cite{Lawrence}  used semidefinite programming to show that the
standard algorithm implements the optimal set of measurements.
All of the above mentioned applications indicate that the method
of SDP is very useful.

Here in this paper,  by using the the convex  semi-definite
programming method,  the robustness of entanglement of  some mixed
entangled quantum states such as: $2\otimes 2$ Bell decomposable
(BD) states, a generic two qubit state in Wootters basis,
iso-concurrence decomposable states, $2\otimes 3$ Bell
decomposable states, $d\otimes d$ Werner and isotropic states, a
one parameter $3\otimes 3$ state and finally multi partite
isotropic state, is calculated exactly, where thus obtained
results are in agreement with those of: $2\otimes 2$ density
matrices, already calculated by one of the authors in\cite{Bell1,
Rob3}. Also an analytic expression is given for separable states
that wipe out all entanglement and it is further shown that they
are on the boundary of separable states as pointed out in
\cite{du}.

The paper is organized as follows:\\ In sections 2 and 3 we give
brief review of semidefinite programming and robustness of
entanglement, respectively. In section 4, by using the
semi-definite programing method we calculate the robustness of
entanglement of some mixed entangled quantum states, such as:
$2\otimes 2$ Bell decomposable (BD)
 states, a generic two qubit state in Wootters basis,
 iso-concurrence decomposable states, $2\otimes 3$ Bell
 decomposable states, $d\otimes d$ Werner and isotropic states, a
 one parameter $3\otimes 3$ state and finally multi partite
 isotropic state. The paper is ended with a brief
conclusion.

\section{Semi-definite programming}\label{semi}
A semidefinite programming(SDP)  is a particular type of convex
optimization problem \cite{optimize}. A semidefinite programming
problem requires minimizing a linear function subject to a linear
matrix inequality (LMI) constraint \cite{optimize1}: \be
\label{sdp1}\begin{array}{cc} \mbox{minimize} & {\cal P}=c^{T}x
\\ \mbox{subject to} & F(x)\geq 0,
\end{array}\ee where c is a given vector,
$x=(x_{1},...,x_{n}), $ and $F(x)=F_{0}+\sum_{i}x_{i}F_{i},$ for
some fixed hermitian matrices $F_{i}$. The inequality sign in
$F(x)\geq 0$ means that $F(x)$ is positive semidefinite.

This problem is called the primal problem. Vectors x that satisfy
the constraint $F(x) \leq 0$ are called primal feasible points,
and if they satisfy $F(x) > 0$ they are called strictly feasible
points. The minimal objective value $c^{T} x$ is by convention
denoted as ${\cal P}^{\ast}$ and is called the primal optimal
value.

The minimization is performed over the vector $x$, whose component
are the variables of the problem. The vector $x$ which satisfies
the LMI, is called a feasible solution, and the set of all
feasible solutions, is called the feasible set.

A very important property of a (SDP) is its convexity, since the
feasible set defined by the above constraints is convex. For this
reason, semidefinite programming has a nice duality structure,
with, the associated dual program being: \be
\label{sdp2}\begin{array}{cc} \mbox{maximize} & -Tr[F_{0}Z] \\
    & Z\geq 0 \\  & Tr[F_{i}Z]=c_{i}. \end{array}\ee

Here the variable is the real symmetric (or Hermitean) matrix Z,
and the data c, $F_{i}$ are the same as in the primal problem.
Correspondingly, matrices Z satisfying the constraints are called
dual feasible (or strictly dual feasible if $Z > 0$). The maximal
objective value $-Tr F_{0}Z$, the dual optimal value, is denoted
as $d^{\ast}$.

The objective value of a primal feasible point is an upper bound
on ${\cal P}^{\ast}$, and the objective value of a dual feasible
point is a lower bound on $d^{\ast}$. The main reason why one is
interested in the dual problem is that one can prove that, under
relatively mild assumptions, ${\cal P}^{\ast} = d^{\ast}$. This
holds, for example, if either the primal problem or the dual
problem are strictly feasible, i.e. there either exist strictly
primal feasible points or strictly dual feasible points. If this
or other conditions are not fulfilled, we still have that
$d^{\ast} \leq {\cal P}^{\ast}$. Furthermore, when both the
primal and dual problem are strictly feasible, one proves the
following optimality condition on x:

A primal feasible $x$ and a dual feasible $Z$ are optimal which
is denoted by $\hat{x}$ and $\hat{Z}$ if and only if  \be
\label{slacknes} F(\hat{x}) \hat{Z}=\hat{Z} F(\hat{x})=0. \ee
This latter condition is called the complementary slackness
condition.

In one way or another, numerical methods for solving SDP problems
always exploit the inequality $d \leq d^{\ast} \leq {\cal
P}^{\ast} \leq {\cal P}$, where d and ${\cal P}$ are the
objective values for any dual feasible point and primal feasible
point, respectively. The difference \be\label{sdp3}{\cal P}-d=
c^{T}x+ Tr[F_{0}Z]= Tr[F(x)Z]\geq 0 \ee is called the duality
gap, and the optimal value ${\cal P}^{\ast}$ is always
"bracketed" inside the interval $[d,{\cal P}]$. These numerical
methods try to minimize the duality gap by subsequently choosing
better feasible points. Under the requirements of the
above-mentioned theorem, the duality gap can be made arbitrarily
small (as far as numerical precision allows).

 Equation (\ref{slacknes}) together with (\ref{sdp2}) and (\ref{sdp1}) constitute a set of necessary and sufficient conditions
for $\hat{x}$ to be an optimal solution to the problem of
(\ref{sdp1}), when both the primal and the dual are strictly
feasible.
\section{Robustness of entanglement}
According to \cite{vidal} for a given entangled state $\rho$ and
separable state $\rho^{\prime\prime}$, a new density matrix
$\rho^{\prime}_{s})$ can be constructed as, \be\label{lrob1}
\rho^{'}_{s}=\frac{1}{s+1}(\rho+s\rho^{\prime\prime}),\quad s\geq
0,
 \ee
 where
it can be either entangled or separable.  It was pointed that
there
 always exits the minimal $\bf s$  corresponding to one $\rho^{\prime\prime}$ such that $\rho^{\prime}_{s}$ is separable. This minimal $s$ is called  the robustness of
 $\rho$ relative to $\rho^{\prime\prime}$, denoted by $R(\rho||\rho^{\prime\prime})$. The
 absolute robustness of $\rho$ is defined as the quantity,
\be\label{lrob2} R(\rho||S)\equiv min
R(\rho||\rho^{\prime\prime}),\quad \rho^{\prime\prime}\in S, \ee
where $S$ is the set of separable states.

 Du et al. in \cite{du} have given a geometrical interpretation
 of robustness and pointed that if $\bf s$ in Eq. (\ref{lrob1}) is
 minimal among all separable states $\rho^{\prime\prime}$, i.e. $\bf s$ is
 the absolute robustness of $\rho$, then $\rho^{\prime\prime}$ and $\rho^{\prime}_{s}$
 in Eq. (\ref{lrob1}) are necessarily on the boundary of the
 separable states.

\section{Robustness of entanglement via  semi-definite programming}
 Unfortunately, the  above mentioned quantity as the most proposed
 measure of entanglement involves extremization which are
 difficult to handle analytically. One of   authors have given analytical
 expression for the robustness of entanglement of some $2\otimes 2$ density matrices
 in \cite{Bell1, Rob3}, here in this section
  we try to obtain robustness of entanglement for many
 categories of states, namely, $2\otimes 2$ Bell decomposable (BD)
 states, a generic two qubit state in Wootters basis,
 iso-concurrence decomposable states, $2\otimes 3$ Bell
 decomposable states, $d\otimes d$ Werner and isotropic states, a
 one parameter $3\otimes 3$ state and finally multi partite
 isotropic state, via semi-definite programming method.
 As we will see in the following, besides the elegance of semi-definite programming in
the  calculation of the robustness of entanglement, also we do not
need  to define any kind of norm for mixed quantum states in
order to calculate their robustness of entanglement as it is done
in Ref. \cite{Bell1, Rob3}.

\subsection{Robustness of entanglement  for Bell-decomposable state}
A Bell decomposable (BD) state is defined by:
\begin{equation}
\rho=\sum_{i=1}^{4}p_{i}\left|\psi_i\right>\left<\psi_i\right|,\quad\quad
0\leq p_i\leq 1,\quad \sum_{i=1}^{4}p_i=1,
 \label{BDS1}
\end{equation}
where $\left|\psi_i\right>$ is Bell state, given by:
\begin{eqnarray}
\label{BS1} \left|\psi_1\right>=\left|\phi^{+}\right>
=\frac{1}{\sqrt{2}}(\left|\uparrow\uparrow\right>
+\left|\downarrow\downarrow\right>), \\
\label{BS2}\left|\psi_2\right>=\left|\phi^{-}\right>
=\frac{1}{\sqrt{2}}(\left|\uparrow\uparrow\right>
-\left|\downarrow\downarrow\right>), \\
\label{BS3}\left|\psi_3\right>=\left|\psi^{+}\right>
=\frac{1}{\sqrt{2}}(\left|\uparrow\downarrow\right>
+\left|\downarrow\uparrow\right>), \\
\label{BS4}\left|\psi_4\right>=\left|\psi^{-}\right>
=\frac{1}{\sqrt{2}}(\left|\uparrow\downarrow\right>
-\left|\downarrow\uparrow\right>).
\end{eqnarray}
In terms of Pauli's matrices, $\rho$ can be written as,

\begin{equation}
\rho=\frac{1}{4}(I\otimes I+\sum_{i=1}^{3}
t_i\sigma_{i}\otimes\sigma_{i}), \label{BDS2}
\end{equation}
where

\begin{equation}\label{t-p}
\begin{array}{rl}
t_1=&p_1-p_2+p_3-p_4,  \\
t_2=&-p_1+p_2+p_3-p_4, \\
t_3=&p_1+p_2-p_3-p_4.
\end{array}
\end{equation}
From the positivity of $\rho$ we get
\begin{equation}\label{T1}
\begin{array}{rl}
1+t_1-t_2+t_3\geq & 0,  \\
1-t_1+t_2+t_3\geq & 0,  \\
1+t_1+t_2-t_3\geq & 0,  \\
1-t_1-t_2-t_3\geq & 0.
\end{array}
\end{equation}
These equations form a tetrahedral  with its vertices located at
$(1,-1,1)$, $(-1,1,1)$, $(1,1,-1)$, $(-1,-1,-1)$ \cite{horo2}. In
fact these vertices denote the Bell states  given in Eqs.
(\ref{BS1}) to (\ref{BS4}), respectively.

On the other hand  $\rho$ given in Eq. (\ref{BDS2}) is separable
if and only if $t_i$ satisfy Eq. (\ref{T1}) and,
\begin{equation}\label{T2}
\begin{array}{rl}
1+t_1+t_2+t_3\geq & 0,  \\
1-t_1-t_2+t_3\geq & 0,  \\
1+t_1-t_2-t_3\geq & 0,  \\
1-t_1+t_2-t_3\geq & 0.
\end{array}
\end{equation}

Inequalities (\ref{T1}) and (\ref{T2}) form an octahedral with its
 vertices located at
$O_1^{\pm}=(\pm 1,0,0)$, $O_2^{\pm}=(0,\pm 1,0)$ and
$O_3^{\pm}=(0,0,\pm 1)$. So, tetrahedral is divided into five
regions. Central regions, defined by octahedral, are separable
states ($p_{k}\leq \frac{1}{2}$). There are also four smaller
equivalent tetrahedral corresponding to entangled states($p_{k}>
\frac{1}{2}$ for only one of $k=1,...,4$), where
$p_{k}=\frac{1}{2}$ denote to boundary between separable and
entangled region. Each tetrahedral takes one Bell state as one of
its vertices. Three other vertices of each tetrahedral form a
triangle which is its common face with octahedral (See Fig. 1).

 Here in this section we evaluate robustness of
entanglement for all BD-states with semi-definite programming
method, and we give an explicit form the corresponding
$\rho^{'}_{s}$ and $\rho^{\prime\prime}$ which are on the
boundary of the separable states.

 Now for a
given BD density matrix \be \label{BDS51}
\rho=\sum_{i=1}^{4}p_{i}\ket{\psi_{i}}\bra{\psi_{i}},\quad
p_{1}>\frac{1}{2},\;\;\sum_{i=2}^{4} p_{i}<\frac{1}{2} \ee and
arbitrary separable density matrix \be \label{BDS4}
\rho^{'}_{s}=\sum_{i=1}^{4}p^{'}_{i}\ket{\psi_{i}}\bra{\psi_{i}}
,\quad\mbox{with}\; p^{'}_{1}\leq\frac{1}{2},\ee according to the
 SDP method explained in  section (\ref{semi}), we have to
optimize ${\cal P}=c^{T}x=-Tr(\Lambda\rho) $ with  \be
\label{BDS5} F(x)=F_{0}+\Lambda
F_{1}=\rho^{'}_{s}+\frac{1}{1+s}(-\rho)\geq0,\ee Therefore, we
have \be {\cal P}=-\Lambda, \mbox{and} \;\;
\Lambda=x=\frac{1}{1+s}\ee Now using the complementary slackness
equation (\ref{slacknes}) with a optimal feasible
pair($\hat{Z},\hat{\Lambda}$), we have \be \label{BDS9}
\hat{Z}(\rho^{'}_{s}-\hat{\Lambda}\rho)=0, \ee or \be
\label{BDS10}
\hat{Z}(I-\hat{\Lambda}\rho\rho^{'^{\dagger}}_{s})=0, \ee where
$\rho^{'^{\dagger}}_{s}$ is pseudo inverse of $\rho^{'}_{s}$.
 By substituting (\ref{BDS51}) and (\ref{BDS4}) into (\ref{BDS10}) and considering the positivity of $\rho^{'}_{s}-\hat{\Lambda}\rho$
 and after some elementary algebra
we arrive at the following results, \be \label{BDS11}
\hat{\Lambda}=min\{\frac{p^{'}_{1}}{p_{1}},\frac{p^{'}_{2}}{p_{2}},\frac{p^{'}_{3}}{p_{3}},\frac{p^{'}_{4}}{p_{4}}
\}. \ee Now defining  \be\label{bdfr}
s\rho^{\prime\prime}=(1+s)\rho^{'}_{s}-\rho, \ee according to
equation (\ref{lrob1})  we get \be s
p^{\prime\prime}_{i}+p_{i}=(1+s) p^{'}_{i},\;\; i=1,2,3,4 \ee
Hence using the above equation we get the following result for the
parameter \be
s=\frac{p_{1}-p^{'}_{1}}{p^{'}_{1}-p^{\prime\prime}}_{1}\ee The
choice $\hat{\Lambda}=\frac{p^{'}_{1}}{p_{1}}$ leads to the rank
three density matrix $\rho^{\prime\prime}$ with
$p^{\prime\prime}_{1}=0$ and the parameter \be
s_{1}=\frac{p_{1}-p^{'}_{1}}{p^{'}_{1}}\ee but other choices of
$\hat{\Lambda}=\frac{p^{'}_{i}}{p_{i}},\;\; i=2,3,4$ leads to
rank three matrices with $p^{\prime\prime}_{1}\neq 0$ and
$s_{i}=\frac{p_{i}-p^{'}_{1}}{p_{1}-p^{\prime\prime}_{1}}.$
Comparing $s_{i}, i=1,2,3,4$, one can show that $s_{1}$ is smaller
than the others, hence for given separable density matrix
$\rho^{'}_{s}$, the choice $\hat{\Lambda}=\frac{p^{'}_{1}}{p_{1}}$
yields the minimum parameter $s_{1}$.

We see that the parameter $s_{1}=\frac{p_{1}}{p^{'}_{1}}-1$ is a
monotonic decreasing function of $p^{'}_{1}$ in the separable
region $0 \leq p^{'}_{1} \leq 1/2$ and its minimum value can be
obtained for $p^{'}_{1}=1/2$ states which lies at the boundary of
separable region.
 Therefore
the robustness of entanglement, that is, the  minimum of
s\be\label{rop2} s=2p_{1}-1\ee corresponds to separable states
$\rho^{'}$ and $\rho^{\prime\prime}$ lying at the corresponding
boundaries $p^{'}_{1}=\frac{1}{2}$ and $p^{\prime\prime}_{1}=0$ of
separable region, in agreement with the results of references
\cite{akhtar1,Bell1,du}.

So far using the SDP optimization method we have proved that for a
given entangled density matrix $\rho$ minimum $s$ in formula
(\ref{lrob1}) is achived for separable states $\rho^{'}_{s}$ and
$\rho^{\prime\prime}$ lying at the boundary of separable region.
One we choose $\rho^{\prime\prime}$ at that part of boundary of
separable region far from $\rho$, one can determine
$\rho^{'}_{s}$ simply from the intersection of  a straight line
drawn from $\rho$ to $\rho^{\prime\prime}$ and the segment of the
boundary of separable region near to $\rho$. As it is shown in
Fig (2) the boundary ${\cal S}^{'}_{1}=P_{1}P_{2}P_{3}$ is near
$\rho$ and others ${\cal S}_{2}=P_{1}P_{3}O^{-}_{2}O^{-}_{3}$,
${\cal S}_{3}=P_{1}P_{2}O^{-}_{1}O^{-}_{2}$, ${\cal
S}_{4}=P_{2}P_{3}O^{-}_{1}O^{-}_{3}$ are far from it. Therefore,
first we have to  choose $\rho^{\prime\prime}$ at one of ${\cal
S}^{'}_{1},{\cal S}_{2},{\cal S}_{3},{\cal S}_{4}$ , but we see
that only the choice of ${\cal S}^{'}_{1}$ leads to
$p^{\prime\prime}_{1}=0$, that is minimum $s_{1}$ and other
choices leads to greater value of parameter $s$. But the choice of
$\rho^{\prime\prime}$ at any point of ${\cal S}^{'}_{1}$=boundary
will yields the same minimum robustness of entanglement, which
result in the choice of $\rho^{'}_{s}$ belong  to a triangle
defined by \be
\begin{array}{cccc} A=\{p^{'}_{1}=\frac{1}{2}, &
p^{'}_{2}=\frac{1}{2}-\frac{1-2p_{2}}{4p_{1}}, &
p^{'}_{3}=\frac{1}{2}-\frac{1-2p_{3}}{4p_{1}}, &
p^{'}_{4}=\frac{1-p_{2}-p_{3}}{2p_{1}}-\frac{1}{2}\}, \\
B=\{p^{'}_{1}=\frac{1}{2}, &
p^{'}_{2}=\frac{1}{2}-\frac{1-2p_{2}}{4p_{1}}, &
p^{'}_{3}=\frac{1-p_{2}-p_{4}}{2p_{1}}-\frac{1}{2}, &
p^{'}_{4}=\frac{1}{2}-\frac{1-2p_{4}}{4p_{1}}\}, \\
 C=\{p^{'}_{1}=\frac{1}{2}, &
p^{'}_{2}=\frac{1-p_{3}-p_{4}}{2p_{1}}-\frac{1}{2}, &
p^{'}_{3}=\frac{1}{2}-\frac{1-2p_{3}}{4p_{1}}, &
p^{'}_{4}=\frac{1}{2}-\frac{1-2p_{4}}{4p_{1}}\}. \end{array} \ee
lying at boundary ${\cal S}^{'}_{1}$.
\subsection{Robustness of entanglement for  $2 \times 2 $ density matrix in Wootters's basis}
Here, we find robustness of a generic two qubit density matrix. To
this aim we first review Wootters's basis as presented by
Wootters in \cite{woot}. Wootters in \cite{woot,} has shown that
for any two qubit density matrix $\rho$ there always exist a
decomposition
\begin{equation}\label{rhox}
\rho=\sum_i\left|x_i\right>\left<x_i\right|,
\end{equation}
called Wootters's basis, such that
\begin{equation}\label{xxt}
\left<x_i|\tilde{x}_j\right>=\lambda_i\delta_{ij},
\end{equation}
where $\lambda_i$ are square roots of eigenvalues, in decreasing
order, of the non-Hermitian matrix $\rho\tilde{\rho}$ and
\begin{equation}\label{rhotilde}
{\tilde \rho}
=(\sigma_y\otimes\sigma_y)\rho^{\ast}(\sigma_y\otimes\sigma_y),
\end{equation}

where $\rho^{\ast}$ is the complex conjugate of $\rho$ when it is
expressed in a standard basis such as
$\{\left|\uparrow\uparrow\right>,
\left|\uparrow\downarrow\right>\},\{\left|\downarrow\uparrow\right>,
\left|\downarrow\downarrow\right>\}$ and $\sigma_y$ represent
Pauli matrix in local basis $\{\left|\uparrow\right>,
\left|\downarrow\right>\}$ . Based on this, the concurrence of the
mixed state $\rho$ is defined by
$\max(0,\lambda_1-\lambda_2-\lambda_3-\lambda_4)$ \cite{woot} to
see the explicit form of the wootters basis of the generic $2\ast
2 $ density matrix see ref. \cite{quant}.

Now let us define states $\left|x^\prime_i \right>$ as
\begin{equation}\label{xprime}
\left|x^\prime_i \right>=\frac{\left|x_i
\right>}{\sqrt{\lambda_i}},\qquad \mbox{for}\,\,i=1,2,3,4.
\end{equation}
Then $\rho$ can be expanded as
\begin{equation}\label{rholambda}
\rho=\sum_i\lambda_i\left|x^\prime_i \right>\left<x^\prime_i
\right|,
\end{equation}
and Eq. (\ref{xxt}) takes the following form
\begin{equation}\label{xpxpt}
\left<x^\prime_i|\tilde{x^\prime_j}\right>=\delta_{ij}.
\end{equation}
Here in this section we obtain the  robustness for a generic two
qubit density matrix with $SDP$ method. Our method of evaluation
of robustness is based on the decomposition of density matrix
given by Wootters in \cite{woot}. By defining $P_i=\lambda_iK_i$,
where $K_i=\left<x^\prime_i|x^\prime_i\right>$, then normalization
condition of $\rho$ leads to
\begin{equation}\label{normal}
Tr(\rho)=\sum_{i=1}^{4}P_i=1, \qquad P_i>0.
\end{equation}
This means that with respect to coordinates $P_i$, the space of
density matrices forms a tetrahedral. With respect to this
representation separability condition
$\lambda_1-\lambda_2-\lambda_3-\lambda_4\leq 0$ takes the
following form
\begin{equation}\label{sepcon}
\frac{P_1}{K_1}-\frac{P_2}{K_2} -\frac{P_3}{K_3}
-\frac{P_4}{K_4}\leq 0.
\end{equation}
The states that saturate inequality (\ref{sepcon}) form a plane
called ${\cal S}_1$ (see Fig. 3). All states violating inequality
(\ref{sepcon}) are entangled states for which $\lambda_1\geq
\lambda_2+\lambda_3+\lambda_4$. These states form an entangled
region with ${\cal S}_1$ as its separable boundary. There exist,
however, three other entangled regions corresponding to  the
dominating  $\lambda_j$ ($j=2,3,4$), respectively. These regions
also define separable planes ${\cal S}_j$. Four planes ${\cal
S}_i$ together with four planes ${\cal S}^{\prime}_i$,
corresponding to $\lambda_i=0$, form an irregular octahedral
corresponding to the separable states. This geometry is similar
to that of Bell decomposable states but here we have an irregular
octahedral associated to separable states \cite{horo}. Figure 3
shows a perspective of this geometry, where two separable planes
${\cal S}_1$ and ${\cal S}^{\prime}_1$ are shown explicitly.

 Now in order to obtain the  robustness of $\rho$, suppose that a ray from $\rho$ is
drawn such that intersects the boundary planes of separable
region at points $\rho^\prime_s$ and $\rho^{\prime\prime}$,
 respectively. Although $\rho^\prime_s$ is necessarily on the plane
${\cal S}_1$, but $\rho^{\prime\prime}$ is allowed to lie on any
plane ${\cal S}^{\prime}_1$, ${\cal S}_2$, ${\cal S}_3$ or ${\cal
S}_4$, where we evaluate robustness for each case separately.

Again for a given generic entangled density matrix $\rho$ in
wootters basis \be \rho=\sum \lambda_{i}
\ket{x^{'}_{i}}\bra{x^{'}_{i}},\;\; \lambda_{1} \geq
\lambda_{2}+\lambda_{3}+\lambda_{4} \ee and an arbitrary separable
density matrix in the same wootters basis \be \label{wootersep}
\rho^{'}_{s}=\sum
\lambda^{'}_{i}\ket{x^{'}_{i}}\bra{x^{'}_{i}},\;\; \lambda^{'}_{1}
< \lambda^{'}_{2}+\lambda^{'}_{3}+\lambda^{'}_{4}  \ee the SDP
optimization of $-Tr(\Lambda \rho)$ with respect to
$\rho^{'}_{s}-\Lambda \rho > 0$ yields \be \label{BDS11}
\hat{\Lambda}=min\{\frac{\lambda_{1}^{'}}{\lambda_{1}},\frac{\lambda_{2}^{'}}{\lambda_{2}},\frac{\lambda_{3}^{'}}{\lambda_{3}},\frac{\lambda_{4}^{'}}{\lambda_{4}}
\}. \ee
 In this case $\rho^{\prime\prime}$ can be written as a convex sum of three
vertices of the plane
\begin{equation}\label{sigma234}
\rho^{\prime\prime}
=\sum_i\lambda^{\prime\prime}_i\left|x^\prime_i\right>\left<x^\prime_i\right|
=a_2\sigma_2+a_3\sigma_3+a_4\sigma_4, \qquad a_2+a_3+a_4=1,
\end{equation}
where $\sigma_i$ are separable states that can be written as
 a convex sum of two corresponding vertices of tetrahedral as
\begin{equation}\label{sigma2}
\sigma_2=\frac{1}{K_3+K_4}\left|x^{\prime}_3\right>\left<x^{\prime}_3\right|
+\frac{1}{K_3+K_4}\left|x^{\prime}_4\right>\left<x^{\prime}_4\right|,
\end{equation}
\begin{equation}\label{sigma3}
\sigma_3=\frac{1}{K_2+K_4}\left|x^{\prime}_2\right>\left<x^{\prime}_2\right|
+\frac{1}{K_2+K_4}\left|x^{\prime}_4\right>\left<x^{\prime}_4\right|,
\end{equation}
\begin{equation}\label{sigma4}
\sigma_4=
\frac{1}{K_2+K_3}\left|x^{\prime}_2\right>\left<x^{\prime}_2\right|
+\frac{1}{K_2+K_3}\left|x^{\prime}_3\right>\left<x^{\prime}_3\right|,
\end{equation}
and $\lambda^{\prime\prime}_i$ are
\begin{equation}\label{lambda1pp}
\lambda^{\prime\prime}_1=0,
\end{equation}
\begin{equation}\label{lambda2pp}
\lambda^{\prime\prime}_2=\frac{a_3}{K_2+K_4}+\frac{a_4}{K_2+K_3}=\frac{\lambda_1}{\lambda_1+\lambda^{'}_1}(\lambda^{'}_2-\frac{\lambda_1}{\lambda^{'}_1}\lambda_{2}),
\end{equation}
\begin{equation}\label{lambda3pp}
\lambda^{\prime\prime}_3=\frac{a_2}{K_3+K_4}+\frac{a_4}{K_2+K_3}=\frac{\lambda_1}{\lambda_1+\lambda^{'}_1}(\lambda^{'}_3-\frac{\lambda_1}{\lambda^{'}_1}\lambda_{3}),
\end{equation}
\begin{equation}\label{lambda4pp}
\lambda^{\prime\prime}_4=\frac{a_2}{K_3+K_4}+\frac{a_3}{K_2+K_4}=\frac{\lambda_1}{\lambda_1+\lambda^{'}_1}(\lambda^{'}_4-\frac{\lambda_1}{\lambda^{'}_1}\lambda_{4}).
\end{equation}
By expanding $\rho^{\prime}_s$ as convex sum of $\rho$ and
$\rho^{\prime\prime}$ and also using the fact that the
coordinates of $\rho^{\prime}_s$ satisfy the equation
\begin{equation}\label{rhopscond}
\frac{P^{\prime}_1}{K_1}-\frac{P^{\prime}_2}{K_2}-\frac{P^{\prime}_3}{K_3}
-\frac{P^{\prime}_4}{K_4}=0,
\end{equation}
after some algebra, coordinates $\lambda^\prime_i$ of
$\rho^\prime_s$ can be written as
\begin{equation}\label{lambda1p}
\lambda^\prime_1=
\frac{\left(\frac{a_2}{K_3+K_4}+\frac{a_3}{K_2+K_4}+\frac{a_4}{K_2+K_3}\right)\lambda_1}
{\frac{a_2}{K_3+K_4}+\frac{a_3}{K_2+K_4}+\frac{a_4}{K_2+K_3}+\frac{C}{2}},
\end{equation}
\begin{equation}\label{lambda2p}
\lambda^\prime_2=
\frac{\left(\frac{a_2}{K_3+K_4}+\frac{a_3}{K_2+K_4}+\frac{a_4}{K_2+K_3}\right)\lambda_2+
\frac{C}{2}\left(\frac{a_3}{K_2+K_4}+\frac{a_4}{K_2+K_3}\right)}
{\frac{a_2}{K_3+K_4}+\frac{a_3}{K_2+K_4}+\frac{a_4}{K_2+K_3}+\frac{C}{2}},
\end{equation}
\begin{equation}\label{lambda3p}
\lambda^\prime_3=
\frac{\left(\frac{a_2}{K_3+K_4}+\frac{a_3}{K_2+K_4}+\frac{a_4}{K_2+K_3}\right)\lambda_3+
\frac{C}{2}\left(\frac{a_2}{K_3+K_4}+\frac{a_4}{K_2+K_3}\right)}
{\frac{a_2}{K_3+K_4}+\frac{a_3}{K_2+K_4}+\frac{a_4}{K_2+K_3}+\frac{C}{2}},
\end{equation}
\begin{equation}\label{lambda4p}
\lambda^\prime_4=
\frac{\left(\frac{a_2}{K_3+K_4}+\frac{a_3}{K_2+K_4}+\frac{a_4}{K_2+K_3}\right)\lambda_4+
\frac{C}{2}\left(\frac{a_2}{K_3+K_4}+\frac{a_3}{K_2+K_4}\right)}
{\frac{a_2}{K_3+K_4}+\frac{a_3}{K_2+K_4}+\frac{a_4}{K_2+K_3}+\frac{C}{2}},
\end{equation}
where $C=\lambda_1-\lambda_2-\lambda_3-\lambda_4$ is the
concurrence of $\rho$. By using the above result one can evaluate
robustness of $\rho$ relative to $\rho^{\prime\prime}$ as
\begin{equation}\label{s1}
s_1=\frac{1}{\hat{\Lambda}}-1=\frac{C}{\frac{2a_2}{K_3+K_4}+\frac{2a_3}{K_2+K_4}+\frac{2a_4}{K_2+K_3}}.
\end{equation}
Analogue to the above method one can evaluate robustness of $\rho$
for the case that $\rho^{\prime\prime}$ lies on the plane ${\cal
S}_2$ i.e. $\hat{\Lambda}=\frac{\lambda^{'}_{2}}{\lambda_{2}}$.
In this case $\rho^{\prime\prime}_s$ can be expanded as convex
sum of three vertices of the plane
\begin{equation}\label{sigma123}
\rho^{\prime\prime}=b_1\sigma_1+b_3\sigma_3+b_4\sigma_4, \qquad
b_1+b_3+b_4=1,
\end{equation}
where
\begin{equation}\label{sigma1}
\sigma_1=\frac{1}{K_1+K_2}\left|x^{\prime}_1\right>\left<x^{\prime}_1\right|
+\frac{1}{K_1+K_2}\left|x^{\prime}_2\right>\left<x^{\prime}_2\right|,
\end{equation}
and $\sigma_3$ and $\sigma_4$ are defined in Eqs. (\ref{sigma3})
and (\ref{sigma4}). Then after some algebra we obtain the
corresponding robustness as
\begin{equation}\label{s2}
s_2=\frac{C}{\frac{2b_3}{K_2+K_4}+\frac{2b_4}{K_2+K_3}}.
\end{equation}
Similarly in cases that separable state $\rho^{\prime\prime}_s$
are on the planes ${\cal S}_3$ and ${\cal S}_4$ we obtain relative
robustness of $\rho$ as
\begin{equation}\label{s3}
s_3=\frac{C}{\frac{2c_2}{K_3+K_4}+\frac{2c_3}{K_2+K_4}},
\end{equation}
and
\begin{equation}\label{s4}
s_4=\frac{C}{\frac{2d_3}{K_2+K_4}+\frac{2d_4}{K_2+K_3}},
\end{equation}
respectively. Equations. (\ref{s1}), (\ref{s2}), (\ref{s3}) and
(\ref{s4}) show that in order to achieve  the minimum robustness
it is enough to consider the case that separable state
$\rho^{\prime\prime}_s$ lies on the plane ${\cal S}^{\prime}_1$.
Hence the choice
$\hat{\Lambda}=\frac{\lambda^{'}_{1}}{\lambda_{1}}$ corresponds
to smaller value of parameter $s_{1}$ than others. Therefore for
the separable density matrix $\rho^{'}_{s}$ given in
(\ref{wootersep}), the minimum value of parameter $s_{1}$, is
given by \be
s_{1}=\frac{\lambda_{1}}{\lambda^{'}_{1}}-1=\frac{p_{1}}{p^{'}_{1}}-1\ee
Obviously $s_{1}$ is a decreasing function of parameter
$p^{'}_{1}$. Again the minimum value of $s_{1}$ can be obtain
from maximum possible value $p^{'}_{1}$, since $s_{1}$ is a
monotonic decreasing function of $p^{'}_{1}$, and its maximum
value corresponds to the separable states with \be
\lambda^{'}_{1}=\lambda^{'}_{2}+\lambda^{'}_{3}+\lambda^{'}_{4},\ee
i.e. , the separable state $\rho^{'}_{s}$  lying on it boundary
of separable region in agreement with \cite{akhtar1,Bell1,du}.
With this consideration we are now allowed to choose coefficients
$a_i$ in such a way that Eq. (\ref{s1}) becomes minimum. It is
easy to see that this happens as long as the coefficient $a_k$
corresponding to the term $\min(K_i+K_j)$ becomes one. Therefore
the robustness of $\rho$ relative to $\rho^{\prime\prime}_s$ is
\begin{equation}\label{rob2}
s=\frac{\min(K_i+K_j)}{2}C
\end{equation}
which is  one of the  main results of  this work. Here the
minimum is taken over all combination of $K_i+K_j$ for
$i,j=2,3,4$. Equation (\ref{rob2}) implies that for two qubit
systems robustness is proportional to the concurrence. We see
that the minimum robustness given in Eq. (2-3) corresponds to
$a_i=\delta_{ik}$, therefore, by using Eq. (\ref{sigma234}) we
get the following result for $\rho^{\prime\prime}$
\begin{equation}\label{rhoppsigma}
\rho^{\prime\prime}_s=\sigma_k.
\end{equation}
Also by using $a_i=\delta_{ik}$ in Eqs. (\ref{lambda1p}) to
(\ref{lambda4p}) one can easily obtain the coordinates
$\lambda^\prime_i$ of separable state $\rho^{\prime}_s$.

As we will show in the next section the Bell decomposable states
correspond to the $K_i=1$ for $i=1,2,3,4$, therefore in Bell
decomposable states Eq. (\ref{rob2}) implies that the robustness
is equal to the concurrence.

One can show that thus obtained robustness is minimum over all
separable states.
\subsection{Iso-concurrence decomposable
states}\label{subsecICD} In this section we define
iso-concurrence decomposable (ICD) states, then we give their
separability condition and evaluate robustness of entanglement.
The iso-concurrence states are defined by
\begin{eqnarray} \label{ICS12}
\left|\phi_1\right>=\cos{\theta}\left|00\right>
+\sin{\theta}\left|11\right>,\qquad
\left|\phi_2\right>=\sin{\theta}\left|00\right>
-\cos{\theta}\left|11\right>, \\
\label{ICS34} \left|\phi_3\right>=\cos{\theta}\left|01\right>
+\sin{\theta}\left|10\right>,\qquad
\left|\phi_4\right>=\sin{\theta}\left|01\right>
-\cos{\theta}\left|10\right>.
\end{eqnarray}
It is quite easy to see that the above states are orthogonal, thus
span the Hilbert space of $2\otimes 2$ systems. Also by choosing
$\theta=\frac{\pi}{4}$  the above states reduce to Bell states.
Now we can define ICD states as
\begin{equation} \label{ICDS}
\rho=\sum_{i=1}^{4}p_{i}\left|\phi_i\right>\left<\phi_i\right|,\quad\quad
0\leq p_i\leq 1,\quad \sum_{i=1}^{4}p_i=1.
\end{equation}
 These states form a four simplex (tetrahedral)  with its
vertices defined by $p_1=1$, $p_2=1$, $p_3=1$ and $p_4=1$,
respectively.

Peres-Horodeckis criterion \cite{peres,horo0} for separability
implies that the state given in Eq. (\ref{ICDS}) is separable if
and only if the following inequalities are satisfied
\begin{eqnarray}
\label{ppt1} (p_1-p_2)\leq
\sqrt{4p_3p_4/\sin^2{2\theta}+(p_3-p_4)^2}, \\ \label{ppt2}
(p_2-p_1)\leq \sqrt{4p_3p_4/\sin^2{2\theta}+(p_3-p_4)^2}, \\
\label{ppt3} (p_3-p_4)\leq
\sqrt{4p_1p_2\sin^2{2\theta}+(p_1-p_2)^2}, \\ \label{ppt4}
(p_4-p_3)\leq \sqrt{4p_1p_2/\sin^2{2\theta}+(p_1-p_2)^2}.
\end{eqnarray}
Inequalities (\ref{ppt1}) to (\ref{ppt4}) divide tetrahedral of
density matrices to five regions.  Central regions, defined by the
above inequalities, form a deformed octahedral and are separable
states. In four other regions one of the above inequality will not
hold, therefore they represent entangled states. Bellow we
consider entangled states corresponding to the violation of
inequality (\ref{ppt1}) i.e. the states which satisfy the
following inequality
\begin{equation}
\label{ICDE1}
(p_1-p_2)>\sqrt{4p_3p_4/\sin^2{2\theta}+(p_3-p_4)^2}.
\end{equation}
In order to obtain the robustness of ICD states, we have to
follow the  method presented by Wootters in \cite{woot}. Starting
from the spectral decomposition for ICD states, and defining
subnormalized orthogonal eigenvectors, the wootters basis of ICD
states can be defined as
$$\ket{x_{1}}=-i\alpha_{1}\sqrt{p_{1}}\ket{\phi_{1}}+i\alpha_{2}\sqrt{p_{2}}\ket{\phi_{2}},$$
$$\ket{x_{2}}=\;\;\;\alpha_{2}\sqrt{p_{1}}\ket{\phi_{1}}+\;\;\;\alpha_{1}\sqrt{p_{2}}\ket{\phi_{2}},$$
\be\ket{x_{3}}=\;\;\;\alpha_{3}\sqrt{p_{3}}\ket{\phi_{3}}-\;\;\;\alpha_{4}\sqrt{p_{4}}\ket{\phi_{4}},\ee
$$\ket{x_{1}}=-i\alpha_{4}\sqrt{p_{3}}\ket{\phi_{3}}-i\alpha_{3}\sqrt{p_{4}}\ket{\phi_{4}},$$
where
$$
\alpha_1=\frac{\left((p_1+p_2)\sin{2\theta}+
\sqrt{4p_1p_2+(p_1-p_2)^2\sin^2{2\theta}}\right)}
{\sqrt{2}\left(4p_1p_2\cos^2{2\theta}+(p_1+p_2)^2\sin^2{2\theta}+
(p_1+p_2)\sin{2\theta}\sqrt{4p_1p_2+(p_1-p_2)^2\sin^2{2\theta}}\right)},
$$
$$
\alpha_2=\frac{\sqrt{2p_1p_2}\cos{2\theta}}
{\left(4p_1p_2\cos^2{2\theta}+(p_1+p_2)^2\sin^2{2\theta}+
(p_1+p_2)\sin{2\theta}\sqrt{4p_1p_2+(p_1-p_2)^2\sin^2{2\theta}}\right)},
$$
\begin{equation}\label{alpha}
\end{equation}
$$
\alpha_3=\frac{\left((p_3+p_4)\sin{2\theta}+
\sqrt{4p_3p_4+(p_3-p_4)^2\sin^2{2\theta}}\right)}
{\sqrt{2}\left(4p_3p_4\cos^2{2\theta}+(p_3+p_4)^2\sin^2{2\theta}+
(p_3+p_4)\sin{2\theta}\sqrt{4p_3p_4+(p_3-p_4)^2\sin^2{2\theta}}\right)},
$$
$$
\alpha_4=\frac{\sqrt{2p_3p_4}\cos{2\theta}}
{\left(4p_3p_4\cos^2{2\theta}+(p_3+p_4)^2\sin^2{2\theta}+
(p_3+p_4)\sin{2\theta}\sqrt{4p_3p_4+(p_3-p_4)^2\sin^2{2\theta}}\right)}.
$$
Now it is easy to evaluate $\lambda_i$ which yields
\begin{equation}\label{lambda1234}
\begin{array}{c}
\lambda_1=\frac{1}{2}\left((p_1-p_2)\sin{2\theta}
+\sqrt{4p_1p_2+(p_1-p_2)^2\sin^2{2\theta}}\right), \\
\lambda_2=\frac{1}{2}\left((p_2-p_1)\sin{2\theta}
+\sqrt{4p_1p_2+(p_1-p_2)^2\sin^2{2\theta}}\right), \\
\lambda_3=\frac{1}{2}\left((p_3-p_4)\sin{2\theta}
+\sqrt{4p_3p_4+(p_3-p_4)^2\sin^2{2\theta}}\right), \\
\lambda_4=\frac{1}{2}\left((p_4-p_3)\sin{2\theta}
+\sqrt{4p_3p_4+(p_3-p_4)^2\sin^2{2\theta}}\right).
\end{array}
\end{equation}
Therefore, coefficients $K_{i}=\frac{p_{i}}{\lambda_{i}}$ are:
\begin{equation}\label{lambda1234}
\begin{array}{c}
K_1=\frac{p_1}{\frac{1}{2}\left((p_1-p_2)\sin{2\theta}
+\sqrt{4p_1p_2+(p_1-p_2)^2\sin^2{2\theta}}\right)}, \\
K_2=\frac{p_2}{\frac{1}{2}\left((p_2-p_1)\sin{2\theta}
+\sqrt{4p_1p_2+(p_1-p_2)^2\sin^2{2\theta}}\right)}, \\
K_3=\frac{p_3}{\frac{1}{2}\left((p_3-p_4)\sin{2\theta}
+\sqrt{4p_3p_4+(p_3-p_4)^2\sin^2{2\theta}}\right)}, \\
K_4=\frac{p_4}{\frac{1}{2}\left((p_4-p_3)\sin{2\theta}
+\sqrt{4p_3p_4+(p_3-p_4)^2\sin^2{2\theta}}\right)}.
\end{array}
\end{equation}
Writing the ICD state in wootters basis, we evaluate its
robustness of entanglement with respect to the set separable
state, diagonal in ICD basis, simply by chossing the separable
states $\rho_{s}^{\prime}$ and $\rho_{s}^{\prime\prime}$ on the
corresponding boundaries, as follows, $$
\rho^{\prime}_{s}=\frac{1}{2}\left((p^{\prime}_1-p^{\prime}_2)\sin{2\theta}
+\sqrt{4p^{\prime}_1p^{\prime}_2+(p^{\prime}_1-p^{\prime}_2)^2\sin^2{2\theta}}\right)\left<x^{\prime}_{1}|x^{\prime}_{1}\right>$$$$+\frac{1}{2}\left((p^{\prime}_2-p^{\prime}_1)\sin{2\theta}
+\sqrt{4p^{\prime}_1p^{\prime}_2+(p^{\prime}_1-p^{\prime}_2)^2\sin^2{2\theta}}\right)\left<x^{\prime}_{2}|x^{\prime}_{2}\right>$$$$+\frac{1}{2}\left((p^{\prime}_3-p^{\prime}_4)\sin{2\theta}
+\sqrt{4p^{\prime}_3p^{\prime}_4+(p^{\prime}_3-p^{\prime}_4)^2\sin^2{2\theta}}\right)\left<x^{\prime}_{3}|x^{\prime}_{3}\right>$$$$+\frac{1}{2}\left((p^{\prime}_4-p^{\prime}_3)\sin{2\theta}
+\sqrt{4p^{\prime}_3p^{\prime}_4+(p^{\prime}_3-p^{\prime}_4)^2\sin^2{2\theta}}\right)\left<x^{\prime}_{4}|x^{\prime}_{4}\right>,$$
\be
\rho^{\prime\prime}_{s}=\frac{1}{2}\left((p^{\prime\prime}_1-p^{\prime\prime}_2)\sin{2\theta}
+\sqrt{4p^{\prime\prime}_1p^{\prime\prime}_2+(p^{\prime\prime}_1-p^{\prime\prime}_2)^2\sin^2{2\theta}}\right)\left<x^{\prime}_{1}|x^{\prime}_{1}\right>\ee$$+\frac{1}{2}\left((p^{\prime\prime}_2-p^{\prime\prime}_1)\sin{2\theta}
+\sqrt{4p^{\prime\prime}_1p^{\prime\prime}_2+(p^{\prime\prime}_1-p^{\prime\prime}_2)^2\sin^2{2\theta}}\right)\left<x^{\prime}_{2}|x^{\prime}_{2}\right>$$$$+\frac{1}{2}\left((p^{\prime\prime}_3-p^{\prime\prime}_4)\sin{2\theta}
+\sqrt{4p^{\prime\prime}_3p^{\prime\prime}_4+(p^{\prime\prime}_3-p^{\prime\prime}_4)^2\sin^2{2\theta}}\right)\left<x^{\prime}_{3}|x^{\prime}_{3}\right>$$$$+\frac{1}{2}\left((p^{\prime\prime}_4-p^{\prime\prime}_3)\sin{2\theta}
+\sqrt{4p^{\prime\prime}_3p^{\prime\prime}_4+(p^{\prime\prime}_3-p^{\prime\prime}_4)^2\sin^2{2\theta}}\right)\left<x^{\prime}_{4}|x^{\prime}_{4}\right>,$$

with the corresponding robustness of entanglement as:
\be\label{rop1} s=\frac{\min(K_i+K_j)}{2}C \ee where $K_{i}$ are
given in (\ref{lambda1234}).

It is obvious that, BD state correspond to particular case of
$\theta=\frac{\pi}{4}$,  ICD ones with $K_i=1$ for $i=1,2,3,4$.
Therefore, the robustness of entanglement given in (\ref{rop1})
become $s=C$ which is agreement with (\ref{rop2})

\subsection{ $2\otimes 3$ Bell decomposable state}
\label{subsecBDS23} In this subsection we obtain robustness of
entanglement
 for Bell decomposable states of $2\otimes 3$ quantum
systems. A Bell decomposable density matrix acting on $2\otimes3$
Hilbert space can be defined by
\begin{equation} \label{BDS23}
\rho=\sum_{i=1}^{6}p_{i}\left|\psi_i\right>\left<\psi_i\right|,\quad\quad
0\leq p_i\leq 1,\quad \sum_{i=1}^{6}p_i=1,
\end{equation}
where $\left|\psi_i\right>$ are Bell states in $H^6\cong
H^2\otimes H^3$ Hilbert space, defined by:

$$
\left|\psi_1\right>=
\frac{1}{\sqrt{2}}(\left|11\right>+\left|22\right>), \qquad
\left|\psi_2\right>=
\frac{1}{\sqrt{2}}(\left|11\right>-\left|22\right>),
$$
\begin{equation}\label{BS123456}
\left|\psi_3\right>=
\frac{1}{\sqrt{2}}(\left|12\right>+\left|23\right>), \qquad
\left|\psi_4\right>=
\frac{1}{\sqrt{2}}(\left|12\right>-\left|23\right>),
\end{equation}
$$ \left|\psi_5\right>=
\frac{1}{\sqrt{2}}(\left|13\right>+\left|21\right>), \qquad
\left|\psi_6\right>=
\frac{1}{\sqrt{2}}(\left|13\right>-\left|21\right>). $$ It is
quite easy to see that the above states are orthogonal and hence
it can  span the Hilbert space of $2\otimes3$ systems. From
Peres-Horodeckis \cite{peres,horo0} criterion for separability we
deduce that the state given in Eq. (\ref{BDS23}) is separable if
and only if the following inequalities are satisfied
\begin{equation}\label{S1}
(p_1-p_2)^2\le(p_3+p_4)(p_5+p_6),
\end{equation}
\begin{equation}\label{S2}
(p_3-p_4)^2\le(p_5+p_6)(p_1+p_2),
\end{equation}
\begin{equation}\label{S3}
(p_5-p_6)^2\le(p_1+p_2)(p_3+p_4).
\end{equation}
In the following we always assume without loss of generality that
$p_1\ge p_2$, $p_3 \ge p_4$ and $p_5 \ge p_6$. Now in order to
obtain robustness of entanglement for BD state given in Eq.
(\ref{BDS23}) we choose $\rho^{\prime}_s=\sum_i
p_i^{\prime}\left|\psi_i\right>\left<\psi_i\right|$ and
$\rho=\sum_i p_i\left|\psi_i\right>\left<\psi_i\right|$. We also
assume without loss of generality that $\rho_s$ lies on the
separable-entangled boundary defined by (all other cases where
$\rho_s$ lies on other surfaces can be treated similarly)
\begin{equation}\label{SS1}
p_1^{\prime}-p_2^{\prime}
=\sqrt{(p_3^{\prime}+p_4^{\prime})(p_5^{\prime}+p_6^{\prime})}.
\end{equation}
Moreover $\rho_s$ must satisfies the other two separability
conditions (\ref{S2}) and (\ref{S3}). This means that entangled
state $\rho$ violates separability condition (\ref{S1}), i.e. we
have
\begin{equation}\label{E1}
p_1\ge p_2 +\sqrt{(p_3+p_4)(p_5+p_6)}.
\end{equation}
Here the boundary of the separable states is given by
\begin{equation}\label{EQ23}\left((p_1-p_2)^2-(p_3+p_4)(p_5+p_6)\right)=0.
\end{equation}

States that saturate inequality (\ref{S1}),(\ref{S2}),(\ref{S3}))
form a plane called ${\cal S}_1$ . All states violating these
three inequality  are entangled states. These states form an
entangled region with ${\cal S}_1$ as its separable boundary.
There exist, however, other entangled regions correspond to the
dominating  $p_j$  ($j=2,..,6$), respectively. These regions also
define separable planes ${\cal S}_j$. The Planes ${\cal S}_i$
together with the planes ${\cal S}^{\prime}_i$, correspond  to
$p_i=0$.

Below in the rest of this subsection we will use Eqs.
(\ref{lrob1}) and (\ref{BDS10}) to calculate robustness of
entanglement for $2\otimes 3$ entangled  Bell decomposable
density matrix  \be \rho=\sum_{i=1}^{6}
p_{i}\ket{\psi_{i}}\bra{\psi_{i}}\;\; with \;\; p_{1} >
p_{2}+\sqrt{(p_{3}+p_{4})(p_{5}+p_{6})},\ee and an arbitrary
separable density matrix \be
\rho^{'}_{s}=\sum_{i=1}^{6}p^{'}_{i}\left|\psi_i\right>\left<\psi_i\right|,\quad\quad
p^{'}_1 \leq p^{'}_2+\sqrt{(p^{'}_3+p^{'}_4)(p^{'}_5+p^{'}_6)}\ee
The SDP optimization of $-Tr(\Lambda\rho)$ with respect to
$\rho^{'}_{s}-\Lambda\rho > 0$ yields \be \label{ICDS4}
\hat{\Lambda}=min\{\frac{p_{1}^{'}}{p_{1}},\frac{p_{2}^{'}}{p_{2}},\frac{p_{3}^{'}}{p_{3}},\frac{p_{4}^{'}}{p_{4}},\frac{p_{5}^{'}}{p_{5}},\frac{p_{6}^{'}}{p_{6}}
\}, \ee The choice of
$\hat{\Lambda}=\frac{p^{\prime}_{i}}{p_{i}}\;\;,i=1,...,6$
consistent with positivity of $\rho - s\rho_{s}^{'}$ implies that
$\rho^{\prime\prime}$ lies at the boundary ${\cal
S}_{i}\;,i=1,...,6$. Numerical calculation indicates that the
minimum s is achieved with the choice of
$\hat{\Lambda}=\frac{p^{\prime}_{1}}{p_{1}}$ (for more details
refer to ref\cite{theses}). Therefore  the robustness of
entanglement becomes \be s_{1}=\frac{p_{1}-p^{'}_{1}}{p^{'}_{1}},
\ee we see that s is a monotonic decreasing function of
$p^{'}_{1}$ and its minimum is achieved for
$p^{'}_{1}=p^{'}_{2}+\sqrt{(p^{'}_{3}+p^{'}_{4})(p^{'}_{5}+p^{'}_{6})}$
which implies that $\rho^{'}_{s}$ lies at the boundary as
separable state, too, in agreement with those results of
ref\cite{theses}.

Now in order to obtain  minimum robustness of entanglement with
respect to the set of separable states diagonal in $2\times 3$, BD
states, all we need to draw a line from $\rho$ to interest the
separable boundary\be
p^{'}_{1}=p^{'}_{2}+\sqrt{(p^{'}_{3}+p^{'}_{4})(p^{'}_{5}+p^{'}_{6})},\ee
at $\rho^{'}_{s}$  and the other boundary  \be \label{ss1}
p^{\prime\prime}_{1}=p^{\prime\prime}_{2}+\sqrt{(p^{\prime\prime}_{3}+p^{\prime\prime}_{4})(p^{\prime\prime}_{5}+p^{\prime\prime}_{6})}.\ee
at $\rho^{\prime\prime}$, respectively.

Therefore, from the relation
$\rho=(1+s)\rho^{'}-s\rho^{\prime\prime}$ we have \be
p_{i}=(1+s)p_{i}^{'}-sp_{i}^{\prime\prime}. \ee
 Hence the robustness of
entanglement $s_{1}$ becomes  \be \label{E13}
s_{1}=\frac{(p_{1}-p_{2})^{2}-(p_{3}+p_{4})(p_{5}+p_{6})}{-2(p_{1}-p_{2})(p^{\prime\prime}_{1}-p^{\prime\prime}_{2})+(p_{3}+p_{4})(p^{\prime\prime}_{5}+p^{\prime\prime}_{6})+(p_{5}+p_{6})(p^{\prime\prime}_{3}+p^{\prime\prime}_{4})}.
\ee where the maximization of the denominator, by using the
Lagrange multipliers method due to existence of constrains
(\ref{ss1}) and normalization of $Tr(\rho^{\prime\prime})=1$,
leads to the following results for the minimums robustness of
entanglement \be \label{smin}
s_{1}=\frac{3((p_{1}-p_{2})^{2}-(p_{3}+p_{4})(p_{5}+p_{6}))}{2(\sqrt{(2p_{1}-1)^{2}+3((p_{1}-p_{2})^{2}-(p_{3}+p_{4})(p_{5}+p_{6}))}-(2p_{2}-1))}.
 \ee

\subsection{Werner states}\label{subsecWerner}
Werner states are the only states that are invariant under local
unitary operations and for $d\otimes d$ systems the Werner states
are defined by \cite{werner}
\begin{equation}\label{Horo4}
\rho=\frac{1}{d^3-d}\left((d-f)I+(df-1){\cal F}\right), \qquad
-1\le f\le 1,
\end{equation}
where $I$ stands for identity operator and ${\cal
F}=\sum_{i,j}\left|i j\right>\left<j i \right|$. It is shown that
Werner state is separable iff $0\le f\le 1$.

Now to obtain  the optimal robustness of entanglement with
respect to accessible separable region, that is, density matrices
of werner type with  $0\le f^{\prime}\le 1$, all we need is to
choose an arbitrary point $\rho^{\prime}_{s}$ in  separable
region $0\leq f^{\prime}\leq 1$. Then the SDP method of
optimization of $-Tr(\rho\Lambda)$ with respect to
$\rho^{\prime}_{s}-\Lambda\rho>0$ yields \be
\hat{\Lambda}=min\{\frac{f^{\prime}+1}{f+1},\frac{1-f^{\prime}}{1-f}\}=\frac{1-f^{\prime}}{1-f},\ee
where the second equality follows from the fact that parameters
$f$ and $f^{\prime}$ are restricted to the regions $f\in (-1,0)$
and $f^{\prime}\in (0,1)$, respectively.

Therefore, for the corresponding parameter $s$ we get \be
s=\frac{f^{\prime}-f}{-f^{\prime}+1}=-1+\frac{1-f}{1-f^{\prime}},\ee
which is a monotonic decreasing function of $f^{\prime}$. Hence,
the optimal robustness of entanglement is $s=-f$ which
corresponds to the  choice  of werner type separability matrix
with $f^{\prime}=0$. On the other hand we have
\be\rho^{\prime\prime}=\rho-(s+1)\rho_{s}^{\prime}=\frac{1}{d(d+1)}\left(I+{\cal
F}\right),\ee which corresponds to the  separable states with
$f=1$ in Eq.(\ref{Horo4}).

Again both separable states $\rho_{s}^{\prime}$ and
$\rho^{\prime\prime}$ are at the boundary of separable region in
agreement with \cite{Bell1,akhtar1,du}.

\subsection{Isotropic states}\label{subsecIsotropic}
The $d\otimes d$ bipartite isotropic states are the only ones that
are invariant under $U\otimes U^\ast$ operations, where $^\ast$
denotes complex conjugation. The isotropic states of $d\otimes d$
systems are defined by \cite{horo3}
\begin{equation}\label{Horo3}
\rho=\frac{1-F}{d^2-1}\left(I-\left|\psi^{+}\right>\left<\psi^{+}\right|\right)
+F\left|\psi^{+}\right>\left<\psi^{+}\right| , \qquad 0\le F\le 1,
\end{equation}
where $\left|\psi^{+}\right>=\frac{1}{\sqrt{d}}\sum_{i}\left|i i
\right>$ is maximally entangled state. It is shown that isotropic
state is separable when $0\le F\le \frac{1}{d}$ \cite{horo3}.

Again to obtain  the optimal robustness of entanglement with
respect to accessible separable region, that is, density matrices
of Isotropic type with  $0\le F^{\prime}\le \frac{1}{d}$, all we
need is to choose an arbitrary point $\rho^{\prime}_{s}$ in
separable region $0\leq F^{\prime}\leq \frac{1}{d}$. Then the SDP
method of optimization of $-Tr(\rho\Lambda)$ with respect to
$\rho^{\prime}_{s}-\Lambda\rho>0$ yields \be
\hat{\Lambda}=min\{\frac{1-F^{\prime}}{1-F},\frac{F^{\prime}}{F}\}=\frac{F^{\prime}}{F},\ee
where the second equality follows from the fact that parameters
$F$ and $F^{\prime}$ are restricted to the regions $F\in (1/d,1)$
and $F^{\prime}\in (0,1/d)$, respectively.

Therefore, for the corresponding parameter $s$ we get \be
s=\frac{F}{F^{\prime}}-1,\ee which is a monotonic decreasing
function of $F^{\prime}$. Hence, the optimal robustness of
entanglement is $s=dF-1$ which corresponds to the choice  of
Isotropic type separability matrix with $F^{\prime}=\frac{1}{d}$.
On the other hand we have
\be\rho^{\prime\prime}=\rho-(s+1)\rho_{s}^{\prime}=\frac{1}{d^{2}-1}\left(I-\left|\psi^{+}\right>\left<\psi^{+}\right|\right),\ee
which corresponds to separable states with $F=0$ in
Eq.(\ref{Horo3}).

Again both separable states $\rho_{s}^{\prime}$ and
$\rho^{\prime\prime}$ are at the boundary of separable region in
agreement with \cite{Bell1,akhtar1,du}.
\subsection{  One parameter $3\otimes 3$ state}\label{subsec33}
Finally let us consider a one parameter state acting on $H^9\cong
H^3\otimes H^3$ Hilbert space as \cite{horo2}
\begin{equation}\label{Horo2}
\rho=\frac{2}{7}\left|\psi^{+}\right>\left<\psi^{+}\right|+\frac{\alpha}{7}\sigma_{+}
+\frac{5-\alpha}{7}\sigma_{-}, \qquad 2\le \alpha \le 5,
\end{equation}
where
\begin{equation}
\begin{array}{l}
\left|\psi^{+}\right>=
\frac{1}{\sqrt{3}}\left(\left|11\right>+\left|22\right>+\left|33\right>\right),
\\
\sigma_{+}=\frac{1}{3}\left(\left|12\right>\left<12\right|
\left|23\right>\left<23\right|+\left|31\right>\left<31\right|\right),
\\
\sigma_{-}=\frac{1}{3}\left(\left|21\right>\left<21\right|
\left|32\right>\left<32\right|+\left|13\right>\left<13\right|\right).
\end{array}
\end{equation}
$\rho$ is separable iff $2 \le \alpha \le 3$, it is bound
entangled iff $3 \le \alpha \le 4$
 and it is distillable entangled state iff $4
\le \alpha \le 5$ \cite{horo2}.

Similarly in order to obtain  the optimal robustness of
entanglement with respect to accessible separable region, that
is, density matrices of $3\otimes 3$ type with  $2\le
\alpha^{\prime}\le 3$, all we need is to choose and arbitrary
point $\rho^{\prime}_{s}$ in separable region $2\leq
\alpha^{\prime} \leq 3$. Then the SDP method of optimization of
$-Tr(\rho\Lambda)$ with respect to
$\rho^{\prime}_{s}-\Lambda\rho>0$ yields \be
\hat{\Lambda}=\frac{\alpha^{\prime}}{\alpha},\ee where the second
equality follows from the fact that parameters $\alpha$ and
$\alpha^{\prime}$ are restricted to the regions $\alpha\in (3,5)$
and $\alpha^{\prime}\in (2,3)$, respectively.

Therefore, for the corresponding parameter $s$ we get \be
s=\frac{\alpha}{\alpha^{\prime}}-1,\ee which is a monotonic
decreasing function of $\alpha^{\prime}$. Hence, the optimal
robustness of entanglement is $s=\frac{\alpha}{3}-1$ which
corresponds to the choice  of $3\otimes 3$ type separability
matrix with $\alpha^{\prime}=3$. On the other hand we have
\be\rho^{\prime\prime}=\rho-(s+1)\rho_{s}^{\prime}=\frac{2}{7}\left|\psi^{+}\right>\left<\psi^{+}\right|+\frac{5}{7}\sigma_{-},\ee
which corresponds to the separable states with $\alpha=0$ in
Eq.(\ref{Horo2}).

Again both separable states $\rho_{s}^{\prime}$ and
$\rho^{\prime\prime}$ are at the boundary of separable region in
agreement with \cite{Bell1,akhtar1,du}.

\subsection{ Multi partite isotropic states}\label{subsecMultiiso}
In this last subsection we obtain robustness of entanglement for
a n-partite d-levels system. Let us consider the following mixture
of completely random state $\rho_0=I/d^n$ and maximally entangled
state $\left|\psi^{+}\right>$
\begin{equation}\label{multirho}
\rho=(1-r)\frac{I}{d^n}+r\left|\psi^{+}\right>\left<\psi^{+}\right|,\qquad
0\le r \le 1,
\end{equation}
where $I$ denotes identity operator in $d^n$-dimensional Hilbert
space and
$\left|\psi^{+}\right>=\frac{1}{\sqrt{d}}\sum_{i=1}^{d}\left|ii\cdot\cdot\cdot
i\right>$. The separability properties of the state
(\ref{multirho}) is considered in Ref. \cite{pitt2}. It is shown
that the above state is separable iff $0\leq r \leq
r_0=\left(1+d^{n-1}\right)^{-1}$.

Again to obtain  the optimal robustness of entanglement with
respect to accessible separable region, that is, density matrices
of multi partite isotropic type with  $0\le r^{\prime}\le r_{0}$,
all we need is to choose an arbitrary point $\rho^{\prime}_{s}$
in separable region $0\leq r^{\prime} \leq r_{0}$. Then the SDP
method of optimization of $-Tr(\rho\Lambda)$ with respect to
$\rho^{\prime}_{s}-\Lambda\rho>0$ yields \be
\hat{\Lambda}=\frac{1+r^{\prime}(d^{n}-1)}{1+r(d^{n}-1)},\ee
where the second equality follows from the fact that parameters
$r^{\prime}$ and $r$ are restricted to the regions $r^{\prime}\in
(0,r_{0})$ and $r\in (r_{0},1)$, respectively.

Therefore, for the corresponding parameter $s$ we get \be
s=\frac{(r-r^{\prime})(d^{n}-1)}{1+r^{\prime}(d^{n}-1)},\ee which
is a monotonic decreasing function of $r^{\prime}$. Hence, the
optimal robustness of entanglement is
$s=\frac{(r-r_{0})(d^{n}-1)}{1+r_{0}(d^{n}-1)}$ which corresponds
to the choice  of multi partite isotropic type separability matrix
with $r^{\prime}=r_{0}$. On the other hand we have
\be\rho^{\prime\prime}=\rho-(s+1)\rho_{s}^{\prime}=\frac{1}{d^{n}-1}(I-\left|\psi^{+}\right>\left<\psi^{+}\right|)
,\ee which corresponds to separable states with
$r=\frac{1}{1-d^n}$ in Eq.(\ref{multirho}).

Again both separable state $\rho_{s}^{\prime}$ and
$\rho^{\prime\prime}$ are at the boundary of separable region in
agreement with \cite{Bell1,akhtar1,du}.

\section{Appendix}In the definition of the robustness we have to minimize
over all separable states. But almost in all of examples, we have
considered  a specific set of  separable states, i.e, the
diagonal separable states (in the basis that the entangled state
itself is diagonal). Here in this appendix we try to show that the
minimum robustness of an entangled diagonal density matrix in a
given basis, with respect to the set of separable diagonal states
in the same basis, is also minimum over the separable sets of
off-diagonal extension of these diagonal separable states. In
\cite{Bell1}, it has been shown that the robustness given in
equation (\ref{BDS4}) is minimum over all Bell decomposable
states. In the following, we generalize it and show that for any
orthogonal basis the off-diagonal elements of $\rho^{\prime}_{s}$
and $\rho^{\prime\prime}$ (in basis that $\rho$ is diagonal) play
no role in  robustness. First we consider the diagonal entangled
density matrices in some orthonormal basis and  we consider the
generic separable states $\rho^{\prime}_{s}$ and
$\rho^{\prime\prime}$ defined by \be \rho^{\prime}_{s}=\sum_{i}
p^{\prime}_{i} \ket{\chi_{i}}\bra{\chi_{i}}+\sum_{i,j}
a_{ij}\ket{\chi_{i}}\bra{\chi_{j}} \ee \be
\rho^{\prime\prime}=\sum_{i} p^{\prime\prime}_{i}
\ket{\chi_{i}}\bra{\chi_{i}}+\sum_{i,j}
b_{ij}\ket{\chi_{i}}\bra{\chi_{j}}, \ee where $a_{ii}=b_{ii}=0$
and $\ket{\chi_{i}}, i=1,2,..$ are  orthogonal states. Then  the
pseudomixture equation (\ref{bdfr}) implies that  the following
equations must hold \be p_{i}=(1+s) p^{\prime}_{i}-s
p^{\prime\prime}_{i},\ee \be\label{minrob1} (1+s) a_{ij}-s
b_{ij}=0. \ee Now one can easily obtain robustness of $\rho$
relative to $\rho^{\prime}_{s}$ as $$ s=\frac{\parallel
\rho-\rho^{\prime}_{s}
\parallel}{\parallel
\rho^{\prime}_{s}-\rho^{\prime\prime}
\parallel}$$ $$=\sqrt{\frac{\sum_{i}(p_{i}-p^{\prime}_{i})^{2}+Tr(A^{2})}{\sum_{i}(p^{\prime}_{i}-p^{\prime\prime}_{i})^{2}+Tr((A-B)^{2})}}$$
\be \label{minrob2}
=\sqrt{\frac{\sum_{i}(p_{i}-p^{\prime}_{i})^{2}+Tr(A^{2})}{\sum_{i}(p^{\prime}_{i}-p^{\prime\prime}_{i})^{2}+\frac{1}{s^{2}}Tr(A^{2})}}\ee
where $A$ and $B$ are Hermitian matrices with non vanishing
off-diagonal  matrix elements $a_{ij}$ and $b_{ij}$, respectively,
and in the last line we have used equation (\ref{minrob1}). By
solving equation (\ref{minrob2}) for robustness $s$ we get
\be\label{minrob3}
s=\sqrt{\frac{\sum_{i}(p_{i}-p^{\prime}_{i})^{2}}{\sum_{i}(p^{\prime}_{i}-p^{\prime\prime}_{i})^{2}}}.\ee
Equation (\ref{minrob2}) shows that off-diagonal elements of
$\rho^{\prime}_{s}$ and $\rho^{\prime\prime}$ (in basis that
$\rho$ is diagonal ) play no role in robustness. Now we consider
the general case of diagonal entangled density matrices  in some
non-orthogonal basis. Obviously in this case the entangled
density matrix $\rho$ can be written as  \be \rho=\sum_{i}
p_{i}\ket{\eta_{i}}\bra{\eta_{i}}=\sum_{i,j}
\tilde{p}_{ij}\ket{\tilde{\eta}_{i}}\bra{\tilde{\eta}_{i}},\ee
where $\ket{\eta_{i}}, i=1,2,...$ are  non-orthogonal states and
$\ket{\tilde{\eta}_{i}}, i=1,2,..$ are  dual of $\ket{\eta_{i}},
i=1,2,...$ , i.e., we have
$\bra{\tilde{\eta}_{i}}\ket{\eta_{j}}=\delta_{ij}, i, j=1,2,..$ .
 Also the separable states $\rho^{\prime}_{s}$ and
$\rho^{\prime\prime}$ can be written as \be
\rho^{\prime}_{s}=\sum_{i} p^{\prime}_{i}
\ket{\eta_{i}}\bra{\eta_{i}}+\sum_{i,j}
a_{ij}\ket{\eta_{i}}\bra{\eta_{j}}= \sum_{i}
\tilde{p^{\prime}}_{i}
\ket{\tilde{\eta}_{i}}\bra{\tilde{\eta}_{i}}+\sum_{i,j}
\tilde{a}_{ij}\ket{\tilde{\eta}_{i}}\bra{\tilde{\eta}_{j}} \ee \be
\rho^{\prime\prime}=\sum_{i} p^{\prime\prime}_{i}
\ket{\eta_{i}}\bra{\eta_{i}}+\sum_{i,j}
b_{ij}\ket{\eta_{i}}\bra{\eta_{j}}= \sum_{i}
\tilde{p^{\prime\prime}}_{i}
\ket{\tilde{\eta}_{i}}\bra{\tilde{\eta}_{i}}+\sum_{i,j}
\tilde{b}_{ij}\ket{\tilde{\eta}_{i}}\bra{\tilde{\eta}_{j}}, \ee
where $a_{ii}=b_{ii}=\tilde{a}_{ii} =\tilde{b}_{ii}=0$. again the
pseudomixture equation (\ref{bdfr}) implies that the following
equations must hold \be p_{i}=(1+s) p^{\prime}_{i}-s
p^{\prime\prime}_{i}, \ee \be 0= (1+s) a_{ij}-s b_{ij}, \ee \be
\tilde{p}_{ii}=(1+s) \tilde{p^{\prime}}_{i} -s
\tilde{p^{\prime\prime}}_{i}, \ee \be \tilde{p}_{ij}= (1+s)
\tilde{a}_{ij}- s \tilde{b}_{ij}.\ee Now the robustness of $\rho$
relative to $\rho^{\prime}_{s}$ can be easily obtained by using
above Equations as $$ s=\frac{\parallel \rho-\rho^{\prime}_{s}
\parallel}{\parallel
\rho^{\prime}_{s}-\rho^{\prime\prime}
\parallel}  $$ \be=\sqrt{\frac{\sum_{i}(p_{i}-p^{\prime}_{i})(\tilde{p}_{ii}-\tilde{p^{\prime}}_{i})}{\sum_{i}(p^{\prime}_{i}-p^{\prime\prime}_{i})(\tilde{p^{\prime}}_{i}-\tilde{p^{\prime\prime}}_{i})}}. \ee
Again this equation  shows that off-diagonal elements of
$\rho^{\prime}_{s}$ and $\rho^{\prime\prime}$ in non-orthogonal
basis  play no role in robustness.
\section{conclusion}
Using the elegant method of convex semidefinite  optimization
method, we have been able to obtain the robustness of some set of
mixed density matrices with respect to some accessible separable
set. In this method we have been able to calculate the robustness
without using any kind of the space of density matrices, where
the results that obtained are in agreement with those of
norm-method of ref\cite{Rob3}.

Also using SDP method we have shown that the separable density
matrices contributing to the robustness lie at the boundary of
accessible separable region.

\newpage

\vspace{10mm}

{\Large {\bf Figure Captions}}

\vspace{10mm}

Figure 1: All BD states are defined as points interior to
tetrahedral. Vertices $P_{1}$, $P_{2}$, $P_{3}$ and $P_{4}$ denote
projectors corresponding to Bell states given in Eqs. (\ref{BS1})
to (\ref{BS4}), respectively. Octahedral corresponds to separable
states.

\vspace{10mm}

Figure 2: Entangled tetrahedral corresponding to singlet state.
Point $t$ denotes a generic state $\rho$ and points $t^\prime$ and
$t^{\prime\prime}$ are on the separable boundary planes.
\vspace{10mm}

Figure 3: The space of a generic two qubit density matrix is
represented by a tetrahedral. Vertices $P_{i}$ for $i=1,2,3,4$
correspond to pure states defined by
$\rho=\lambda_i\left|x^{\prime}_i\right>\left<x^{\prime}_i\right|$.
Irregular octahedral corresponds to separable states. Separable
planes ${\cal S}_1$ and ${\cal S}^\prime_1$ are shown explicitly.


\begin{thebibliography}{99}
\bibitem{Bell1} S. J.Acktarshenas, M. A. Jafarizadeh Europ. J. Phys. D  v.25, N.3.
293,
(2003).
\bibitem{Rob3} S. J. Acktarshenas, M. A. Jafarizadeh,Robustness of entanglement
 for two qubit density matrix, quant-ph/0211156.
\bibitem{du}  J. F. Du, M. J. Shi, X. Y. Zhou and R. D. Han, Phys. Lett. A, 267, pp 244-250, (2000).
\bibitem{ben1} C. H. Bennett, G. Brassard,
C. Cr\'{e}peau, R. jozsa, A Peres and W. K. Wootters,  Phys. Rev.
Lett.  70, 1895, (1993).
\bibitem{ben2} C. H. Bennett, and S. J. Wiesner,
 Phys. Rev. Lett.  69, 2881, (1992).
\bibitem{ben3}
 C. H. Bennett, D. P. DiVincenzo, J. A. Smolin and W.K.
Wootters,  Phys. Rev. A  54, 3824, (1996).
\bibitem{ved1} V. Vedral, M. B. Pienio, M. A. Rippin and P. L.
Knight.
 Phys. Rev. Lett.  78, 2275, (1995).
\bibitem{ved2} V. Vedral and M. B. Plenio,
  Phys. Rev. A  57, 1619, (1998).
\bibitem{woot} W. K. Wootters,  Phys. Rev. Lett.
 80, 2245, (1998).
\bibitem{vidal} G. Vidal and R. Tarrach,
  Phys. Rev. A  59, 141, (1999).
\bibitem{vidwer} G. Vidal, R.F. Werner.
 Phys. Rev. A 65, 032314, (2002).
\bibitem{rudolph} O. Rudolph. quant-ph/0202121.
\bibitem{Rajagopal} A. K. Rajagopal, and R. W. Rendell.  Phys. Rev. A 65, 032328 (2002).
\bibitem{Simon} C. Simon, J. Kempe. Phys. Rev. A, Vol. 65 (5), 052327 (2002).
\bibitem{Luo04} M. X. Goemans and D. P. Williamson.  Journal of the Association for Computing Machinery,
42(6): 1115-1145, (1995).
\bibitem{Luo03} Z. Q. Luo.  Mathematical Programming
Series B, 97: 177-207, (2003).
\bibitem{Luo02} T. N. Davidson, Z.-Q. Luo, and K. M. Wong. IEEE
Transactions on Signal Processing, 48(5): 1433-1445, (2000).
\bibitem{Luo01} W. K. Ma, T. N. Davidson, K. M. Wong, Z.-Q. Luo, and P. C. Ching.  IEEE
Transactions on Signal Processing, 50: 912-922, (2002).
\bibitem{Luo05} S. Boyd, L. El Ghaoui, E. Feron, and V. Balakrishnan. Society for
Industrial and Applied Mathematics, (1994).
\bibitem{Doherty} A. C. Doherty, P. A. Parrilo, and F. M.
Spedalieri. Phys. Rev. Lett. 88, 187904, (2002),
\bibitem{Parrilo} P. A. Parrilo, A. C. Doherty, and F. M. Spedalieri.  Proceedings of the 41st IEEE Conference of Decision
and Control, (2002).
\bibitem{13} Alexei Kitaev. Quantum coin flipping. Talk at Quantum Information Processing 2003, slides and
video at http://www.msri.org, December (2002).
\bibitem{17} E. M. Rains.  IEEE Transactions on Information
Theory, 47(7): 2921–2933, November (2001).
\bibitem{1} H. Barnum, M. Saks, and M. Szegedy.  In Proceedings of the 18th IEEE Annual Conference on
Computational Complexity, pages 179–193, (2003).
\bibitem{10} C. W. Helstrom. Quantum Detection and Estimation Theory. Academic Press, (1976).
\bibitem{3} Y. C. Eldar, A. Megretski, and G. C. Verghese.  IEEE Transactions on Information
Theory, 49(4):1007–1012, (2003).
\bibitem{Lawrence} Lawrence.  Ip,  Shor's Algorithm is Optimal,   University of California,
Berkeley, USA MU Seminar,  http://www.qcaustralia.org (2003).
\bibitem{optimize} L. Vandenberghe, S. Boyd  SIAM Review  38,
49 - 95, (1996).
\bibitem{optimize1} L. Vandenberghe and S. Boyd,
http://www.stanford.edu/~boyd/cvxbook.html (unpublished).
\bibitem{horo2} P. Horodecki, M. Horodecki and R. Horodecki,
 Phys. Rev. Lett.  82, 1056, (1999).
\bibitem{akhtar1} S. J. Akhtarshenas, M. A. Jafarizadeh, J. Phys. A: Math. Gen. 37,   2965-2982, (2004)
\bibitem{horo} R. Horodecki, and M. Horodecki,
 Phys. Rev. A  54, 1838, (1996).
\bibitem{quant} S. J. Akhtarshenas, M. A. Jafarizadeh Quantum Inf.
Comput., 3 229-248, (2003).
\bibitem{peres} A. Peres,  Phys. Rev. Lett.  77, 1413, (1996).
\bibitem{horo0} M. Horodecki, P. Horodecki and R. Horodecki,
 Phys. Lett. A   223, 1, (1996).
\bibitem{theses} S. J. Akhtarshenas,
 Investigation of Quantum correlation in pure and mixed
multipartite systems, Ph.D Thesis, Tabriz University,   (2003).
\bibitem{werner} R. F. Werner,
 Phys. Rev. A  40, 4277, (1989).
\bibitem{horo3} M. Horodecki and P. Horodecki,
 Phys. Rev. A  59, 4206, (1999).
\bibitem{pitt2} A. O. Pittenger, and M.H. Rubin,
 Optics Communications       179,  447, (2000).
\end{thebibliography}
\end{document}